\documentclass[twocolumn,aps,pra,showpacs,superscriptaddress]{revtex4}
\usepackage[usenames,dvipsnames]{color}

\usepackage{graphicx}
\usepackage{amsmath,amssymb}
\usepackage{color}
\usepackage{graphics}
\usepackage{epsfig}

\newcommand{\be}{\begin{equation}}
\newcommand{\ee}{\end{equation}}
\newcommand{\eea}{\end{eqnarray}}
\newcommand{\bea}{\begin{eqnarray}}

\newcommand{\mean}[1]{\ensuremath{\langle{#1}\rangle}}

\newcommand{\eins}{\openone}

\newcommand{\HH}{\ensuremath{\mathcal{H}}}

\newcommand{\XX}{\ensuremath{\mathcal{X}}}
\newcommand{\SCAL}{\ensuremath{\mathcal{S}}}

\newcommand{\ket}[1]{\ensuremath{|#1\rangle}}

\newcommand{\bra}[1]{\ensuremath{\langle#1|}}

\newcommand{\kommentar}[1]{}
\newcommand{\trace}{{\rm tr}}

\renewcommand{\vr}{\ensuremath{\varrho}}

\newcommand{\forget}[1]{}

\begin{document}


\title{Compatibility and noncontextuality for sequential measurements}

\author{Otfried G{\"u}hne}
 \affiliation{Institut f\"ur Quantenoptik und Quanteninformation,
 \"Osterreichische Akademie der Wissenschaften, Technikerstr.~21A, A-6020 Innsbruck, Austria}
 \affiliation{Institut f\"ur Theoretische Physik, Universit\"at
 Innsbruck, Technikerstr.~25, A-6020 Innsbruck, Austria}
\author{Matthias Kleinmann}
\affiliation{Institut f\"ur Quantenoptik und Quanteninformation,
\"Osterreichische Akademie der Wissenschaften, Technikerstr.~21A, A-6020 Innsbruck, Austria}

\author{Ad\'{a}n Cabello}
\affiliation{Departamento de F\'{\i}sica Aplicada II,
 Universidad de Sevilla, E-41012 Sevilla, Spain}
\author{Jan-{\AA}ke Larsson}
\affiliation{Institutionen f\"or Systemteknik och Matematiska
  Institutionen, Link\"opings Universitet, SE-581 83 Link\"oping,
  Sweden}
\author{Gerhard~Kirchmair}
 \affiliation{Institut f\"ur Quantenoptik und Quanteninformation,
 \"Osterreichische Akademie der Wissenschaften, Technikerstr.~21A, A-6020 Innsbruck, Austria}
 \affiliation{Institut f{\"u}r Experimentalphysik, Universit{\"a}t
 Innsbruck, Technikerstr.~25, A-6020 Innsbruck, Austria}
\author{Florian Z\"ahringer}
 \affiliation{Institut f\"ur Quantenoptik und
 Quanteninformation, \"Osterreichische Akademie der
 Wissenschaften, Technikerstr.~21A, A-6020 Innsbruck, Austria}
 \affiliation{Institut f{\"u}r Experimentalphysik,
 Universit{\"a}t Innsbruck, Technikerstr.~25, A-6020 Innsbruck,
 Austria}
\author{Rene Gerritsma}
 \affiliation{Institut f\"ur Quantenoptik und
 Quanteninformation, \"Osterreichische Akademie der
 Wissenschaften, Technikerstr.~21A, A-6020 Innsbruck, Austria}
 \affiliation{Institut f{\"u}r Experimentalphysik,
 Universit{\"a}t Innsbruck, Technikerstr.~25, A-6020 Innsbruck,
 Austria}
 \author{Christian~F.~Roos}
 \affiliation{Institut f\"ur
 Quantenoptik und Quanteninformation, \"Osterreichische Akademie
 der Wissenschaften, Technikerstr.~21A, A-6020 Innsbruck, Austria}
 \affiliation{Institut f{\"u}r Experimentalphysik,
 Universit{\"a}t Innsbruck, Technikerstr.~25, A-6020 Innsbruck,
 Austria}

\date{\today}


\begin{abstract}
A basic assumption behind the inequalities used for testing noncontextual 
hidden variable models is that the observables measured on the same 
individual system are perfectly compatible. However, compatibility is not 
perfect in actual experiments using sequential measurements. We discuss 
the resulting ``compatibility loophole'' and present several methods 
to rule out certain hidden variable models which obey a kind of extended 
noncontextuality. Finally, we present a detailed analysis of experimental 
imperfections in a recent trapped ion experiment and apply our analysis 
to that case.
\end{abstract}



\maketitle


\section{Introduction}
\label{Sec1}


Since the early days of quantum mechanics (QM), it has been debated 
whether or not QM can be completed with additional hidden variables 
(HVs), which would eventually account for the apparent indeterminism 
of the results of single measurements in QM, and may end into a more 
detailed deterministic description of the world 
\cite{VonNeumann31, EPR35, Bohr35}.
The problem of distinguishing QM from HV theories, however, cannot be addressed 
unless one makes additional assumptions about the structure of the HV 
theories. 
Otherwise, for a given experiment, one can just take the observed probability 
distributions as a HV model \cite{WW01}. Moreover, there are explicit HV 
theories, such as Bohmian mechanics \cite{BH93, Holland93}, which can 
reproduce all experiments up to date.

In the 1960s, it was found out that HV models reproducing the predictions 
of QM should have some peculiar and highly nonclassical properties. The most famous result in this
direction is Bell's theorem \cite{Bell64}. Bell's theorem
states that local HV models cannot reproduce the quantum
mechanical correlations between local measurements on some entangled
states. In principle, the theorem just states a conflict
between two descriptions of the world: QM and local HV models.
However, the proof of Bell's theorem by means of an inequality
involving correlations between measurements on distant systems,
which is satisfied by any local HV model, but is violated by
some quantum predictions \cite{CHSH69}, allows us to take a
step further and test whether or not the world itself can be
described by local HV models \cite{ADR82, WJSWZ98, RKMSIMW01,
MMMOM08, RWVHKCZW09}. More recently, a similar approach has
been used to test whether or not the world can be reproduced
with some specific nonlocal HV models \cite{Leggett03,
GPKBZAZ07, BBGKLLS08}.

A second seminal result on HV models reproducing QM is the
Kochen-Specker (KS) theorem \cite{Specker60, Bell66, KS67}. To
motivate it, one first needs the notion of compatible
measurements: two or more measurements are compatible, if they
can be measured jointly on the same individual system without
disturbing each other (i.e., without altering their results).
Compatible measurements can be made simultaneously or in any
order, and can be repeated any number of times on the same
individual system and always must give the same result
independently of the initial state of the system.

Second, one needs the notion of noncontextuality.
A context is a set of compatible measurements. A physical
model is called noncontextual if it assigns to a measurement a
result independently of which other compatible measurements are
carried out. There are some scenarios where the assumption of
noncontextuality is specially plausible. For instance, in the
case of measurements on distant systems, or in the case that
the measurements concern different degrees of freedom of the
same system and the degrees of freedom can be accessed
independently.

In a nutshell, the KS theorem states that noncontextual HV
models cannot reproduce QM. This impossibility occurs already
for a single three-level system, so it is not related to
entanglement.

There have been several proposals to test the KS theorem
\cite{RS93, CG98, SZWZ00, SBZ01, Larsson02}, but there also
have been debates whether the KS theorem can be experimentally
tested at all \cite{Meyer99, Kent99, Mermin99, CK00, HKSS01,
Appleby02, Cabello02, Breuer02a, Breuer02b, BK04}.
Nevertheless, first experiments have been performed, but these
experiments required some additional assumptions \cite{MWZ00,
HLZPG03, HLBBR06, BKSSCRH09}. Furthermore, the notion of
contextuality has been extended to state preparations
\cite{Spekkens05} and experimentally investigated
\cite{SBKTP09}.

Quite recently, several inequalities have been proposed which
hold for all noncontextual models, but are violated in QM,
potentially allowing for a direct test \cite{CFRH08,
KCBS08, Cabello08, BBCP09}. A remarkable feature of some 
noncontextuality inequalities is that the violation is 
independent of the quantum state of the system \cite{Cabello08, BBCP09}. 
In this paper, we will call these inequalities KS inequalities, since
the proof of the KS theorem in Ref.~\cite{KS67} is also valid for any quantum 
state of the system. Very recently, several experiments have found violations of
noncontextual inequalities \cite{KZGKGCBR09, BKSSCRH09, ARBC09,
LHGSZLG09, MRCL09}. Three of these experiments have found
violations of a KS inequality for different states
\cite{KZGKGCBR09, ARBC09} or for a single (maximally mixed)
state \cite{MRCL09}. In these experiments, compatible
observables are measured sequentially.

The measurements in any experiment are never perfect. In tests
of {noncontextuality} inequalities, these imperfections can be
interpreted as a failure of the assumption that the observables
measured sequentially on the same system are perfectly
compatible. What if this compatibility is not perfect? We will
refer to this problem as the ``compatibility loophole''. 
The main aim of this paper is to give a detailed discussion 
of this loophole and demonstrate that, despite of this loophole, 
still classes of HV models which obey a generalized definition 
of noncontextuality can be experimentally ruled out.

The paper is organized as follows: In Sec.~\ref{Sec2} we give
precise definitions of compatibility and noncontextuality,
focusing on the case of sequential measurements. We also review
some inequalities which have been proposed to test
noncontextual HV models.

In Sec.~\ref{Sec3} we discuss the case of not perfectly
compatible observables. We first derive an inequality which
holds for any HV model, however, this inequality is not
experimentally testable. Then, we consider several possible
extensions of noncontextuality. By that, we mean replacing our
initial assumption of noncontextuality for perfectly compatible
observables by a new one, which covers also nearly compatible
observables and implies the usual noncontextuality if the
measurements are perfectly compatible. We then present several
experimentally testable inequalities which hold for HV models
with some generalized version of noncontextuality, but which
are violated in QM. One of these inequalities has already been
found to be violated in an experiment \cite{KZGKGCBR09}. In
Sec.~\ref{Sec4} we present details of this experiment.

In Sec.~\ref{Sec5} we present two explicit contextual HV models
which violate all investigated inequalities. These models,
which do not satisfy the assumptions of extended
noncontextuality, are useful to understand which
counterintuitive properties a HV model must have to reproduce
the quantum predictions. Other contextual HV models for
{contextuality} experiments have been proposed in
Ref.~\cite{La Cour09a}. Finally, in Sec.~\ref{Sec6},
 we conclude and discuss consequences of our work for future experiments.


\section{Hidden variable models and noncontextuality}
\label{Sec2}



\subsection{Joint or sequential measurements}


In the scenario originally used for discussing noncontextuality
\cite{KS67}, a measurement device is treated as a {single}
device producing outcomes for several compatible measurements
{(i.e., a context)}. When treating the measurement device
in this manner, the whole context is needed to produce any
output at all. In this joint measurement, one of the settings 
of the measurement device is always specifically associated with 
one of the outcomes, in the sense that another measurement 
device exists that takes only that setting as input and 
gives an identical outcome as output. 
This is checked by repeatedly making a joint measurement and the
corresponding compatible single measurements in any possible order.
This is at the basis of the noncontextuality argument. 
The argument goes: {\it precisely because} another context-less device
   exists that can measure the outcome of interest, there is good
   reason to assume that this outcome is independent of the context in
   the joint measurement.

In this paper we discuss sequential individual measurements,
rather than joint measurements. It might be argued that the
version of the noncontextuality assumption needed in this
scenario is more {restrictive} on the HV model than the
version used for joint measurements. This would mean that a
test using a sequential setup would be weaker than a test using
a joint measurement setup, because it would rule out fewer HV
models. 
However, the motivation for assuming non-contextuality even in the
joint measurement setup is the existence of the individual 
measurements and their compatibility and repeatability when 
combined with joint context-needing measurements. Therefore, 
the assumptions needed in the sequential measurements setting 
are equally well-motivated as the assumptions needed in the joint
measurement setting.

In fact, the sequential setting is closer to the actual
motivation of assuming noncontextuality: there exist individual
context-less measurement devices that give the same results as
the joint measurements, \emph{and we actually use them in
experiment}. Furthermore, from an experimental point of view, a
changed context in the joint measurement device corresponds to
a physically entirely different setup even for the unchanged
setting within the context, so it is difficult to maintain that
the outcome for the unchanged setting is unchanged from
physical principles {\cite{Bell66, Bell82}}.
Motivating physically unchanged outcomes is much easier in the sequential
setup, since the device used is physically identical for the
unchanged setting.

Therefore, in this paper we consider the situation where
sequences of measurements are made on an individual physical
system. Throughout the paper, we consider only dichotomic
measurements with outcomes $\pm1$, but the results can be
generalized to arbitrary measurements. The question is: under
which conditions can the results of such measurements be
explained by a HV model? More precisely, we ask which
conditions a HV model has to violate in order to reproduce the
quantum predictions.


\subsection{Notation}


The following notation will be used in the discussed HV models:
$\lambda$ is the HV, drawn with a distribution $p(\lambda)$
from a set $\Lambda$. The distribution summarizes all
information about the past, including all preparation steps and
all measurements already performed. Causality is assumed, so
the distribution is independent of any event in the future. It
rather determines all the probabilities of the results of all
possible future sequences of measurements. We assume that, for
a fixed value of the HV, the outcomes of future sequences of
measurements are deterministic, hence all indeterministic
behavior stems from the probability distribution. This is
similar to the investigation of Bell inequalities, where any
stochastic HV model can be mapped onto a deterministic one
where the HV is not known \cite{WW01, Peres99}.

In an experiment, one first prepares a ``state'' via certain
preparation procedures (which may include measurements). One
always regards a state preparation as a procedure which can be
repeated. At the HV level, it will therefore lead to an
experimentally accessible probability distribution $p_{\rm
exp}(\lambda).$ The HV model hence enables the experimenter to
repeatedly prepare the same distribution. In a single instance
of an experiment, one obtains a state determined by a single
value $\lambda$ of the HV. The probability for this instance is
distributed according to the distribution $p_{\rm
exp}(\lambda),$ and reflects the inability of the experimenter
to control which particular value of the HVs has been prepared
in a single {instance}.

Continuing, we denote by $A_i$ the measurement of the
{observable (or measurement device)} $A$ at the position 
$i$ in the sequence. For example, $A_1B_2C_3$ denotes the 
sequence of measuring $A$
first, then $B$, and finally $C$. An outcome from a
measurement, e.g., $B_2$ from the above sequence, is denoted
$v(B_2|A_1 B_2 C_3)$. The product of three outcomes is denoted
$v(A_1B_2C_3)=v(A_1|A_1B_2C_3)v(B_2|A_1B_2C_3)v(C_3|A_1B_2C_3).$
Given a probability distribution $p(\lambda)$, we write
probabilities $p(B_2^+|A_1 B_2 C_3)$ [or $p(B_2^+ C_3^-|A_1 B_2
C_3)$] for the probability of obtaining the value $B_2=+1$ (and
$C_3=-1$) when the sequence $A_1 B_2 C_3$ is measured. One can
also consider mean values like
$\mean{B_2|A_1B_2C_3}=p(B_2^+|A_1B_2C_3)-p(B_2^-|A_1B_2C_3)$,
or the mean value of the complete sequence, $\mean{A_1 B_2
C_3}= p[v(A_1 B_2 C_3)=+1] - p[v(A_1 B_2 C_3)=-1].$


\subsection{Compatibility of measurements}
\label{def_of_compatibility}

In the simplest case, compatibility is a relation between a
pair of measurements, $A$ and $B.$ For that, let $\SCAL_{AB}$
denote the (infinite) set of all sequences, which use only
measurements of $A$ and $B,$ that is, $\SCAL_{AB}=\{A_1, B_1, A_1
A_2,A_1 B_2, B_1 A_2,\ldots\}$. Then, we formulate:

{\it Definition~1.---}Two observables $A$ and $B$ are
compatible if the following two conditions are fulfilled:

(i) For any instance of a state (i.e. for any $\lambda$) and for any sequence $S \in \SCAL_{AB},$ the
obtained values of $A$ and $B$ remain the same,
\begin{subequations}
\begin{align}
 &v(A_k | S) = v(A_l | S), 
\\
&v(B_m | S) = v(B_n | S), 
\end{align}
\end{subequations}
where $k,l,m,n$ are all possible indices for which the
considered observable is measured at the positions $k,l,m,n$ in
the sequence $S$. [Equivalently, we could require that $p(A_k^+
A_l^-|S)= 0$, etc., for all preparations corresponding to some
$p_\mathrm{exp}(\lambda)$.]

(ii) For any state preparation [i.e., for any $p_{\rm
exp}(\lambda)$], the mean values of $A$ and $B$ during the measurement
of any two sequences $S_1, S_2 \in \SCAL_{AB}$ are equal,
\begin{subequations}
\begin{align}
 &\mean{A_k|S_1}=\mean{A_l|S_2},
\\
 &\mean{B_m|S_1}=\mean{B_n|S_2}.
\end{align}
\end{subequations}
Clearly, conditions (i) and (ii) are necessary conditions for
compatible observables, in the sense that two observables which
violate any of them cannot reasonably called compatible.

It is important to note that the compatibility of two observables 
is experimentally testable by repeatedly preparing all possible 
$p_{\rm exp}(\lambda)$. The fact that this set is infinite is not 
a specific problem here, as any measurement device or physical 
law can only be tested in a finite number of cases. A crucial 
point in a HV model is that the set of all experimentally accessible 
probability distributions $p_{\rm exp}(\lambda)$ might not coincide 
with the set of all possible distributions $p(\lambda)$. We will 
discuss this issue in Sec.~\ref{seccondpd}.

It should be noted that the conditions (i) and (ii) are not
minimal, cf.\ Appendix~A for a discussion. In particular, we
emphasize that (ii) does not necessarily follow from (i), as we
illustrate by the following example: Consider a HV model where,
for any $\lambda$, all $v(A_k | S)$ are $+1$ when the first
measurement in $S$ is $A_1$, while they are $-1$ when the first
measurement is $B_1$. The values $v(B_m | S)$ are always $+1.$
Then, condition (i) is fulfilled, while (ii) is violated, since
$\mean{A}=1$ but $\mean{A_2|B_1A_2}=-1$.

Let us compare our definition of compatibility to the notion of
``equivalent measurements'' introduced by Spekkens in
Ref.~\cite{Spekkens05}. In this reference, two measurements are
called equivalent if, for any state preparation, the
probability distributions of the measurement outcomes for both
measurements are the same. This is similar to our condition
(ii), but disregards repeated measurements on individual systems as in (i).
Interestingly, using this notion and POVMs, one can prove the
contextuality of a quantum-mechanical two-level system
\cite{Spekkens05}.

Finally, it should be added that the notion of compatibility 
is extended in a straightforward manner to three or more observables. 
For instance, if three observables $A,B,C$ are investigated, one 
considers the set $\SCAL_{ABC}$ of all
measurement sequences involving measurements of $A,B,$ or $C$
and extends the conditions (i) and (ii) in an obvious way. This
is equivalent to requiring the pairwise compatibility of
$A,B,C$, cf.\ Appendix~A.


\subsection{Definition of noncontextuality for sequential measurements}


Noncontextuality means that the value of any observable $A$ 
does not depend on which other compatible observables are measured 
jointly with $A$. For our models, we formulate noncontextuality 
as a condition on a HV model as follows:

{\it Definition 2.---}Let $A$ and $B$ be observables in a HV
model, where $A$ is compatible with $B$. We say that the 
HV model is noncontextual if it assigns, for any $\lambda$, 
an outcome of $A$ which is independent of whether $B$ is 
measured before or after $A,$ that is,
\begin{equation}
v(A_1) = v(A_2|B_1 A_2).
\label{ncdef}
\end{equation}
Hence, for these sequences we can write down $v(A)$ as 
being independent of the sequence. If the condition is 
not fulfilled, we call the model contextual.

It is important to note that the condition (\ref{ncdef})
is an assumption about the model and --- contrary to the 
definition of compatibility --- not experimentally testable. 
This is due to the fact that for a given instance of a 
state (corresponding to some unknown $\lambda$) the experimenter
has to decide whether to measure $A$ or $B$ first.

{From} this definition and the time ordering, it follows 
immediately that, if $A$ is compatible with $B$  and $A$ 
is also compatible with $C,$ then for noncontextual models
\begin{equation}
v(A_1|A_1 B_2)= v(A_2|B_1 A_2) = v(A_1|A_1 C_2) = v(A_2|C_1 A_2).
\label{ncdef123}
\end{equation}
holds. This is the often used definition of noncontextual models, 
stating that the value of $A$ does not depend on whether $B$ or 
$C$ is measured  before, jointly with, or after it.

This definition can directly be extended to three or more 
compatible observables. For instance, if $\{A,B,C\}$ are 
compatible, then noncontextuality means that for any $\lambda$,
\begin{align}
 v(A_1) &= v(A_2 | B_1A_2) = v(A_2 | C_1A_2)
 \nonumber \\
&= v(A_3 | B_1B_2A_3) = v(A_3 | B_1C_2A_3)
 \nonumber \\
&= v(A_3 | C_1B_2A_3) = v(A_3 | C_1C_2A_3).
\end{align}
Of course, the equalities in the second and third 
line follow, if the first line holds for any $\lambda$ 
and the HV model allows to see the measurement of $B_1$
or $C_1$ as a preparation step. Again, if $\{A,a,\alpha\}$ 
is another set of compatible observables, one can derive 
consequences similar to Eq.~(\ref{ncdef123}).


\subsection{Inequalities for noncontextual HV models}


Here we will discuss several previously introduced inequalities
involving compatible measurements, which hold for any
noncontextual HV model, but which are violated for certain
states and observables in QM. Later, these inequalities are
extended to the case where the observables are not perfectly
compatible.


\subsubsection{CHSH-like inequality}


To derive a first inequality, consider the mean value
\begin{equation}
\mean{\chi_{\rm CHSH}} = \mean{A B} + \mean{B C} + \mean{C D} -
\mean{D A}.
\label{CHSHpre}
\end{equation}
If the measurements in each average are compatible [i.e., the
pairs $(A,B)$, $(B,C)$, $(C,D)$, and $(D,A)$ are compatible
observables)], then a noncontextual HV model has to assign a
fixed value to each measurement, and the model predicts
\begin{equation}
 |\mean{\chi_{\rm CHSH}}| \leq 2.
 \label{chshineq}
\end{equation}
In QM, on a two-qubit system, one can take the observables
\begin{eqnarray}
A= \sigma_x \otimes \openone, & \;\;\;\; & B = \openone \otimes
\frac{(\sigma_z + \sigma_x)} {\sqrt{2}}, \nonumber
\\
C= \sigma_z \otimes \openone, & \;\;\;\; & D = \openone \otimes
\frac{(\sigma_z - \sigma_x)} {\sqrt{2}}, \label{chshobservables}
\end{eqnarray}
then, the measurements in each sequence are commuting and hence
compatible, but the state
\begin{equation}
\ket{\phi^+}=(\ket{00}+\ket{11})/\sqrt{2}
\end{equation}
leads to a value of $\mean{\chi_{\rm CHSH}} = 2\sqrt{2},$
therefore not allowing any noncontextual description. The
choice of the observables in Eq.~(\ref{chshobservables}) 
is, however, by no means unique, if one transforms all of 
them via the same global unitary transformation, another set 
is obtained, and the state leading to the maximal violation 
does not need to be entangled. In fact, the two-qubit notation 
is only chosen for convenience and could be replaced by a 
formulation with a single party using a four-level system. 
For example, if we take the observables
\begin{eqnarray}
A= \sigma_x \otimes \sigma_x, & \;\;\;\; &
B = \frac{1}{\sqrt{2}}
 \left({\begin{matrix}
 1 & 1 & 0 & 0 \\
 1 & -1 & 0 & 0 \\
 0 & 0 & -1 & 1 \\
 0 & 0 & 1 & 1
 \end{matrix}}\right), \nonumber \\
C= \sigma_z \otimes \openone, & \;\;\;\; &
D = \frac{1}{\sqrt{2}}
 \left({\begin{matrix}
 1 & -1 & 0 & 0 \\
 -1 & -1 & 0 & 0 \\
 0 & 0 & -1 & -1 \\
 0 & 0 & -1 & 1
 \end{matrix}}\right), \label{chshobservables02}
\end{eqnarray}
then, the measurements in each sequence are commuting and hence
compatible, but the product state
\begin{equation}
 \ket{\Psi}=\ket{x^+}\ket{0}=(\ket{00}+\ket{10})/\sqrt{2}
\end{equation}
leads to a value of $\mean{\chi_{\rm CHSH}} = 2\sqrt{2},$
therefore not allowing any noncontextual description.


\subsubsection{The KCBS inequality}


As a second inequality, we take the pentagram inequality introduced
by Klyachko, Can, Binicio\u{g}lu, and Shumovsky (KCBS)
\cite{KCBS08}. Here, one takes five dichotomic observables and
considers
\begin{equation}
\mean{\chi_{\rm KCBS}} = \mean{A B} +\mean{B C} +\mean{C D} +\mean{D
E} + \mean{E A}. \label{kcbsineq}
\end{equation}
If the observables in each mean value are compatible and
noncontextuality is assumed, it can be seen that
\begin{equation}
 \mean{\chi_{\rm KCBS}} \geq -3
 \label{KCBSinequality}
\end{equation}
holds. However, using appropriate measurements on a three-level system,
there are
qutrit states which give a value of $\mean{\chi_{\rm KCBS}} =
5- 4 \sqrt{5} \approx -3.94,$ also leading to contradiction
with noncontextuality.


\subsubsection{An inequality from the Mermin-Peres square}


For the third inequality, we take the one introduced in
Ref.~\cite{Cabello08}. Consider the mean value
\begin{align}
 \mean{\chi_{\rm KS}} = & \mean{A B C}+ \mean{a b c} +
 \mean{\alpha \beta \gamma} + \mean{A a \alpha} + \mean{B b \beta}
 \nonumber \\
  & - \mean{C c \gamma}.
\label{ksinequalityq}
\end{align}
If the measurements in each expectation value are compatible,
then any noncontextual HV model has to assign fixed values to
each of the nine occurring measurements. Then, one can see that
\begin{equation}
 \mean{\chi_{\rm KS}} \leq 4. \label{mpineq}
\end{equation}
However, on a two-qubit system, one can choose the observables of
the Mermin-Peres square \cite{Peres90, Mermin90}
\begin{equation}
\begin{array}{ccc}
A=\sigma_z \otimes \openone, &\;\;\;\; 
B= \openone \otimes \sigma_z, &\;\;\;\; 
C= \sigma_z \otimes \sigma_z,\\
a= \openone \otimes \sigma_x, &\;\;\;\; 
b= \sigma_x \otimes \openone, &\;\;\;\; 
c= \sigma_x \otimes \sigma_x,\\
\alpha = \sigma_z\otimes\sigma_x ,&\;\;\;\; 
\beta = \sigma_x \otimes\sigma_z ,&\;\;\;\; 
\gamma = \sigma_y \otimes\sigma_y.
\end{array}
\label{mpsquare}
\end{equation}
The observables in any row or column commute and are therefore
compatible. Moreover, the product of the observables in any row
or column equals $\openone$, apart from the last column, where
it equals $-\openone.$ Hence, for any quantum state,
\begin{equation}
\mean{\chi_{\rm KS}} = 6
\end{equation}
holds. The remarkable fact in this result is that it shows that
any quantum state reveals nonclassical properties if the
measurements are chosen appropriately.


\section{Not perfectly compatible measurements}
\label{Sec3}


In any real experiment, the measurements will not be perfectly
compatible. Hence, the notion of noncontextuality does not
directly apply. The experimental violation of inequalities like
(\ref{chshineq}), (\ref{KCBSinequality}), and (\ref{mpineq})
proves that one cannot assign to the measurement devices
independent outcomes $\pm 1$. However, a model that is not
trivially in conflict with QM also has to explain the
measurement results of sequences of incompatible observables,
such as e.g.\ the results from measuring $A_1C_2$ for the
observables of the CHSH-like inequality. Therefore, it 
is not straightforward to find out which are the implications 
of these violations on the structure of the possible HV models.
The reason is that the assumption that incompatible measurements 
have predetermined independent outcomes is not physically
plausible.

To deal with this problem, we will derive extended versions of
the inequalities (\ref{chshineq}), (\ref{KCBSinequality}), and
(\ref{mpineq}), which are valid even in the case of imperfect
compatibility. We will first derive an inequality which is an
extension of inequality~(\ref{chshineq}) and which holds for
any HV model. This inequality, however, contains terms which
are not experimentally accessible. Then, we investigate how
these terms can be connected to experimental quantities, if
certain assumptions about the HV model are made. We will
present three types of testable inequalities, the first two
start from condition (i) of Definition~1, while the third one
uses condition (ii).

First we consider nearly compatible observables. We show that,
if the observables fulfill the condition (i) of Definition~1 to
some extent and if assumptions about the dynamics of
probabilities in a HV model are made, then  these HV
models can be experimentally refuted.

In the second approach, we consider the case that a certain
finite number of compatibility tests has been made. For some
runs of the experiment the tests are successful [i.e., no error
occurs when checking condition (i)], and in some runs errors
occur. We then assume that the subset of HVs, where
noncontextuality holds is at least as large as the subset where
the compatibility tests are successful. We then show that HV
models of this type can, in principle, be refuted
experimentally.

Finally, in the third approach, we also consider assumptions
about the possible distributions $p_{\rm exp}(\lambda),$ and
show that if the condition (ii) of Definition~1 is nearly
fulfilled, then again this type of HV models can experimentally
be ruled out.

We will discuss these approaches using the CHSH-like inequality
(\ref{chshineq}). At the end of the section, we will also
explain how the inequalities (\ref{KCBSinequality}) and
(\ref{mpineq}) have to be modified, in order to test these
different types of HV models.


\subsection{CHSH-like inequality for all HV models}


To start, consider a HV model with a probability distribution
$p(\lambda)$ and let $p[(A_1^+|A_1) \mbox{ and }(B_1^+|B_1)]$
denote the probability of finding $A^+$ if $A$ is measured
first and $B^+$ if $B$ is measured first. This probability is
well defined in all HV models of the considered type, but it is
impossible to measure it directly, as one has to decide whether
one measures $A$ or $B$ first. Our aim is now to connect it to
probabilities arising in sequential measurements, as this will
allow us to find contradictions between HV models and QM.

First, note that
\begin{align}
 p[(A_1^+| & A_1) \mbox{ and }(B_1^+|B_1)]
 \leq
 p[A_1^+, B_2^+| A_1 B_2]+
\nonumber
\\
&
+
p[(B_1^+|B_1) \mbox{ and } (B_2^-|A_1 B_2)].
\label{bound1}
\end{align}
This inequality is valid because if $\lambda$ is such that it
contributes to $p[(A_1^+| A_1) \mbox{ and } (B_1^+| B_1)],$ then
either the value of $B$ stays the same when measuring $A_1 B_2$
(hence $\lambda$ contributes to $p[A_1^+, B_2^+| A_1 B_2]$)
or the value of $B$ is flipped and $\lambda$ contributes to
$p[(B_1^+| B_1) \mbox{ and } (B_2^-| A_1 B_2)].$ The first
term $p[A_1^+, B_2^+| A_1 B_2]$ is directly measurable
as a sequence, but the second term is not experimentally
accessible.

Let us rewrite
\begin{eqnarray}
\mean{AB} &=& 1 - 2 p[(A_1^+| A_1) \mbox{ and }(B_1^-|B_1)]
\nonumber
\\
&&
 - 2 p[(A_1^-| A_1) \mbox{ and }(B_1^+|B_1)],
\end{eqnarray}
as the mean value obtained from the probabilities $p[(A_1^\pm| A_1)
\mbox{ and }(B_1^\pm|B_1)].$ Then, using Eq.~(\ref{bound1}), it
follows that
\begin{equation}
\mean{A_1 B_2} - 2 p^{\rm flip}[AB] \leq \mean{AB} \leq
\mean{A_1 B_2} + 2 p^{\rm flip}[AB],
\end{equation}
where we used $ p^{\rm flip}[AB] =p[(B_1^+|B_1) \mbox{ and }
(B_2^-|A_1 B_2)]+ p[(B_1^-|B_1) \mbox{ and } (B_2^+|A_1 B_2)].$
This $p^{\rm flip}[AB]$ can be interpreted as a probability that
$A$ flips a predetermined value of $B$.

Furthermore, using Eqs.~(\ref{CHSHpre}) and (\ref{chshineq}), we obtain
\begin{align}
|\mean{\XX_{\rm CHSH}} |\leq 2 (1+p^{\rm flip}[AB]+p^{\rm flip}[CB]
 \nonumber \\
 +p^{\rm flip}[CD]+p^{\rm flip}[AD]),
 \label{chshwithnoise}
\end{align}
where
\begin{equation}
\mean{\XX_{\rm CHSH}} := \mean{A_1 B_2} + \mean{C_1 B_2} + \mean{C_1
D_2} - \mean{A_1 D_2}.
\end{equation}
Inequality (\ref{chshwithnoise}) holds for any HV model and is
the generalization of inequality~(\ref{chshineq}). Note that
for perfectly compatible observables, the flip terms in
inequality~(\ref{chshwithnoise}) vanish if the assumption of
noncontextuality is made. Then, this results in
inequality~(\ref{chshineq}).


\subsection{First approach: Constraints on the disturbance and the dynamics of the HV}


The terms $p^{\rm flip}[AB]$, etc.\ in
inequality~(\ref{chshwithnoise}) are not experimentally
accessible. Now we will discuss how they can be experimentally
estimated when some assumptions on the HV model are made.

In order to obtain an experimentally testable version of
inequality~(\ref{chshwithnoise}), we will assume that
\begin{align}
p[(B_1^+|& B_1) \mbox{ and } (B_2^-| A_1 B_2)]
\nonumber
\\
&
\leq
p[(B_1^+| B_1) \mbox{ and }
(B_1^+,B_3^-| B_1 A_2 B_3)]
\nonumber
\\
&
\equiv
p[B_1^+,B_3^-| B_1 A_2 B_3].
\label{assumption}
\end{align}
This assumption is motivated by the experimental procedure: Let
us assume that one has a physical state, for which one surely
finds $B_1^+$ if $B_1$ is measured first, but finds $B_2^-$ if
the sequence $A_1 B_2$ is measured. Physically, one would
explain this behavior as a disturbance of the system due to the
experimental procedures when measuring $A_1$. The left-hand
side of Eq.~(\ref{assumption}) can be viewed as the amount of
this disturbance. The right-hand side quantifies the
disturbance of $B$ when the sequence $B_1 A_2 B_3$ is measured.
In real experiments, it can be expected that this disturbance
is larger than when measuring $A_1 B_2$, because of the
additional experimental procedures involved. Note that in real
experiments, a measurement of $B$ will also disturb the value
of $B$ itself, as can be seen from the fact that sometimes the
values of $B_1$ and $B_2$ will not coincide, if the sequence
$B_1 B_2$ is measured.

It should be stressed, however, that we do not assume that the set of HV values
giving $[(B_1^+| B_1) \mbox{ and } (B_2^-| A_1 B_2)]$ is
contained in the set giving $(B_1^+,B_3^-| B_1 A_2 B_3)$, the
assumption only relates the sizes of these two sets.

In addition, by a similar reasoning, the assumption~(\ref{assumption}) 
may be relaxed to 
\be
p[(B_1^+|B_1) \mbox{ and } (B_2^-| A_1 B_2)] \leq p[B_1^+,B_k^-| B_1 S
A_{k-1} B_k],
\label{assumption2}
\ee
where $S$ is a given finite sequence of measurements from
$\SCAL_{AB}.$ Again, if the measurements are nearly compatible,
this type of HV models can be ruled out experimentally.

Assumption (\ref{assumption}) gives an measurable upper bound
to $p^{\rm flip}[AB].$ One directly has
\begin{align}
|\mean{\XX_{\rm CHSH}}| \le & 2 (1
 + p^{\rm err}[B_1 A_2 B_3]
 + p^{\rm err}[B_1 C_2 B_3] \nonumber
\\
& + p^{\rm err}[D_1 C_2 D_3]
 + p^{\rm err}[D_1 A_2 D_3]),
\label{chshwithnoise2}
\end{align}
where we used
\be
p^{\rm err}[B_1 A_2 B_3]
= p[B_1^+,B_3^-| B_1 A_2 B_3] + p[B_1^-,B_3^+| B_1 A_2 B_3],
\label{perrdef}
\ee
denoting the total disturbance probability of $B$ when measuring
$B_1 A_2 B_3.$

The point of this inequality is that if the observable pairs
$(A,B)$, $(C,B)$, $(C,D)$, and $(A,D)$ fulfill approximately
the condition (i) in the definition of compatibility, the terms
$p^{\rm err}$ will become small, and a violation of
inequality~(\ref{chshwithnoise2}) can be observed. In
Ref.~\cite{KZGKGCBR09} it was found that $\mean{\XX_{\rm CHSH}}
- 2 ( p^{\rm err}[B_1 A_2 B_3]+ p^{\rm err}[B_1 C_2 B_3] +
p^{\rm err}[D_1 C_2 D_3]+ p^{\rm err}[D_1 A_2 D_3])= 2.23(5)$.
Hence this experiment cannot be described by HV models which
fulfill Eq.~(\ref{assumption}), see also Section IV.


\subsection{Second approach: Assuming noncontextuality for the 
set of HVs where the observables are compatible}


Let us discuss a different approach to obtain experimentally
testable inequalities. For that, consider the case that the
experimenter has measured a (finite) set of sequences in
$\SCAL_{AB}$ in order to test the validity of condition (i) in
the definition of compatibility. He finds that the conditions
are violated or fulfilled with certain probabilities. In terms
of the HV model, there is a certain subset $\Lambda_{AB}
\subset \Lambda$ of all HVs where {\it all} tests in the finite
set of experimentally performed compatibility tests succeed and
through the observed probabilities the experimenter can
estimate the volume of this set.

In this situation, one can assume that, for each HV $\lambda
\in \Lambda_{AB}$ (where all the measured compatibility
requirements are fulfilled), the assumption of noncontextuality
is also valid. More precisely, one can assume that $v(A_1|A_1
B_2)= v(A_2|B_1 A_2)$ in Eq.~(\ref{ncdef}) holds for all
$\lambda \in \Lambda_{AB}$. One may support this assumption if
one considers noncontextuality as a general property of nature,
since this is the usual noncontextuality assumption for the HV
model where the HVs are restricted to $\Lambda_{AB}.$

To see that this assumption leads to an experimentally testable
inequality, consider the case where the experimenter has tested
all sequences up to length three, that is all sequences from
$\SCAL_{AB}^{(3)} = \{A_1A_2A_3, A_1A_2B_3, \ldots, B_1B_2B_3\}$
and has determined, for each of them, the probability $p^{\rm
err}(S)$ that some measurement, which is performed two or three
times in the sequence is disturbed. For sequences like $B_1 A_2
B_3$, this is exactly $p^{\rm err}[B_1 A_2 B_3]$ defined in
Eq.~(\ref{perrdef}). However, now we have additional error
terms like $p^{\rm err}[B_1 B_2 A_3] = p[B_1^+,B_2^-| B_1 B_2
A_3] + p[B_1^-,B_2^+| B_1 B_2 A_3]$ and $p^{\rm err}[B_1 B_2
B_3] = 1 - p[B_1^+,B_2^+ B_3^+| B_1 B_2 B_3] - p[B_1^-,B_2^-
B_3^-| B_1 B_2 B_3],$ etc. These probabilities are not
completely independent: due to the time ordering, a $\lambda$
that contributes to $p^{\rm err}[B_1 B_2 A_3]$ (or $p^{\rm
err}[A_1 A_2 B_3]$) will also contribute to $p^{\rm err}[B_1
B_2 B_3]$ (or $p^{\rm err}[A_1 A_2 A_3]$). Consequently,
relations like $p^{\rm err}[B_1 B_2 A_3] \leq p^{\rm err}[B_1
B_2 B_3]$ hold.

Let us define
\begin{equation}
 p^{\rm err}[\SCAL_{AB}^{(3)}]
= \Big(\!\!\!\!\sum_{S \in \SCAL_{AB}^{(3)}}p^{\rm err}[S]\Big) 
- p^{\rm err}[B_1 B_2 A_3] - p^{\rm err}[A_1 A_2 B_3].
\end{equation}
Here, we have excluded two $p^{\rm err}$ in the sum, as the
$\lambda$'s which contribute to them are already counted in
other terms. With this definition, for a given distribution
$p_{\rm exp}(\lambda)$, a lower bound to the probability
of finding a $\lambda$ where condition (i) from Definition~1 is
fulfilled, is
\begin{equation}
p[\Lambda_{AB}] \geq 1- p^{\rm err}[\SCAL_{AB}^{(3)}].
\end{equation}
{From} that and the assumption that $v(A_1|A_1 B_2)= v(A_2|B_1
A_2)$ on $\Lambda_{AB}$, it directly follows that
\begin{equation}
 p^{\rm flip}[AB] \leq p^{\rm err}(\SCAL_{AB}^{(3)}),
\end{equation}
giving a measurable upper bound to $p^{\rm flip}[AB].$ Finally,
the experimentally testable inequality
\begin{align}
\label{stochcontext}
 |\mean{\XX_{\rm CHSH}}| & \le 2 (1
 +p^{\rm err}[\SCAL_{AB}^{(3)}]
 +p^{\rm err}[\SCAL_{CB}^{(3)}]
 \nonumber \\
 & +p^{\rm err}[\SCAL_{CD}^{(3)}]
 +p^{\rm err}[\SCAL_{AD}^{(3)}])
\end{align}
holds. This inequality is similar to
inequality~(\ref{chshwithnoise2}), but it contains more error
terms. Nevertheless, a violation of this inequality in ion-trap
experiments might be feasible in the near future (see
Sec.~\ref{Sec4}).

This result deserves two further comments. First, in the
derivation we assumed a pointwise relation; namely, for all
$\lambda \in \Lambda_{AB}$, the noncontextuality assumption
$v(A_1|A_1 B_2)= v(A_2|B_1 A_2)$ holds. Of course, we could
relax this assumption by assuming only that the volume of the
set where $v(A_1|A_1 B_2)= v(A_2|B_1 A_2)$ holds is not smaller
than the volume of $\Lambda_{AB}.$ Under this condition,
Eq.~(\ref{stochcontext}) still holds.

Second, when comparing the second approach with the first one,
one finds that the first one is indeed a special case of the
second one. In fact, from a mathematical point of view,
the first approach is the same as the second one, if in the
second approach only the compatibility test $S=B_1 A_2 B_3$ is
performed. Consequently, inequality~(\ref{chshwithnoise2}) is
weaker than (\ref{stochcontext}). However, note that the first
approach came from a different physical motivation. Further,
assuming a pointwise relation for the first approach is very
assailable, as only one compatibility test is made. But, as we
have seen, a relation between the volumes suffices. A pointwise
relation can only be motivated if all experimentally feasible
compatibility tests are performed.


\subsection{Third approach: Certain probability distributions cannot
  be prepared}
\label{seccondpd}


The physical motivation of the third approach is as follows:
The experimenter can prepare different probability
distributions $p_{\rm exp}(\lambda)$ and check their properties.
For instance, he can test to which extent the condition (ii)
{in Definition~1} is fulfilled. However, in a general HV
model there might be probability distributions $p(\lambda)$
that do not belong to the set of experimentally accessible
$p_{\rm exp}(\lambda).$ One might be tempted to believe that
this difference is negligible and that the properties that can
be verified for the $p_{\rm exp}(\lambda)$ hold also for some
of the $p(\lambda).$ In this approach we will show that this
belief can be experimentally falsified. More specifically, we
show that if only four conditional probability distributions
have the same properties as all $p_{\rm exp}(\lambda),$ then
a contradiction with QM occurs.

So let us assume that the experimenter has checked that the
observables $A$ and $B$ fulfill condition (ii) in Definition~1
approximately. He has found that
\begin{equation}
|\mean{B_1|B_1 A_2}-\mean{B_2|A_1 B_2}| \leq \varepsilon_{AB}
\end{equation}
for all possible (or, at least, a large number of) $p_{\rm
exp}(\lambda).$ This means that, for experimentally accessible
distributions $p_{\rm exp}(\lambda)$, one has that
\begin{align}
|p(B_1^+|B_1 A_2)-p(B_2^+|A_1 B_2)| &\leq \varepsilon_{AB}/2,
\nonumber
\\
|p(B_1^-|B_1 A_2)-p(B_2^-|A_1 B_2)| &\leq \varepsilon_{AB}/2,
\label{probabilitybound}
\end{align}
as can be seen by direct calculation.

Let us consider the flip probability $ p^{\rm flip}[AB] =
p[(B_1^+| B_1) \mbox{ and } (B_2^-| A_1 B_2)] + p[(B_1^-| B_1)
\mbox{ and } (B_2^+| A_1 B_2)] $ again. Here, the probability
$p$ stems from the initial probability distribution
$p(\lambda).$ One can consider the conditional probability
distributions $q^\pm(\lambda)$ which arise from $p(\lambda)$ if
the result of $B_1$ is known. Physically, the conditional
distributions describe the situation for an observer, who knows
that the experimenter has prepared $p(\lambda)$ but has the
additional information that measurement of $B_1$ will give $+1$
or $-1.$ With that, we can rewrite
\begin{align}
p^{\rm flip}[AB]
&=
q^+(B_2^-| A_1 B_2) p(B_1^+| B_1)
\nonumber
\\
&
+
q^-(B_2^+| A_1 B_2) p(B_1^-| B_1).
\end{align}
Now let us assume that these conditional probability
distributions have the same properties as all accessible
distributions $p_{\rm exp}(\lambda).$ Then, the bounds in
Eq.~(\ref{probabilitybound}) also have to hold for $q^\pm$.
Since $ p(B_1^+| B_1)+ p(B_1^-| B_1)=1,$ it follows directly
that $ p^{\rm flip}[AB] \leq \varepsilon_{AB}/2. $ Hence, under
the assumption that some conditional probability
distributions in the HV model have similar properties as the
preparable $p_{\rm exp}(\lambda),$ the inequality
\begin{equation}
\mean{\XX_{\rm CHSH}} \leq 2 + \varepsilon_{AB}+\varepsilon_{CB} +
\varepsilon_{CD}+\varepsilon_{AD} \label{chshwithnoise3}
\end{equation}
holds. A violation of it implies that, in a possible HV model,
certain conditional probability distributions have to be
fundamentally different from experimentally preparable
distributions.

Again, this result deserves some comments. First, note that the
tested bound in Eq.~(\ref{probabilitybound}) does not have to
hold for all probability distributions in the theory. In an
experiment testing Eq.~(\ref{chshwithnoise3}) with some
$\hat{p}_{\rm exp}(\lambda)$ only assumptions about four
conditional probability distributions (corresponding to two
possible second measurements with two outcomes) have to be
made. In fact, assuming Eq.~(\ref{probabilitybound}) for
$\delta$-distributions (i.e., a fixed HV $\lambda$) is not very
physical, as in this case the left-hand side of these equations
is 0 or 1.

Second, finding an experimental violation of
Eq.~(\ref{chshwithnoise3}) shows that these four distributions
have properties significantly different from all preparable
$p_{\text{exp}}(\lambda).$ In other words, one may conclude
that in a possible HV model describing such an experiment, it
must be forbidden to prepare $\hat{p}_{\rm exp}(\lambda)$ with
additional information about the result of $B$ or $D.$

To make this last point more clear, consider the situation
where the experimenter has prepared $\hat{p}_{\rm exp}(\lambda)$ 
and a second physicist has the additional knowledge that the result 
of $B_1$ will be $+1$, if it would be measured as 
a first instance. Both physicists disagree on the probability distribution 
$\hat{p}_{\text{exp}}$ and $q^+$, but that is not the central 
problem because this occurs in any classical model as well. 
The point is that $q^+$ cannot be prepared: If the
experimenter measures $B_1$ and keeps only the cases where he
finds $+1$ he obtains a new experimentally accessible
probability distribution $\tilde{p}_{\text{exp}}$. But this
will not be the same as the probability distribution $q^+$,
because in this case, the first measurement has already been
made.


\subsection{Application to the KCBS inequality and the 
KS inequality (\ref{mpineq})}
\label{subsec:KSineqnonperf}


In the previous discussion, we used the CHSH like inequality
(\ref{chshineq}) to develop our ideas. Clearly, one could also
start from inequalities (\ref{KCBSinequality}) and
(\ref{mpineq}) to obtain testable inequalities for the types of
HV models discussed above.

For the KCBS inequality (\ref{kcbsineq}) this can be done with the same
methods as before, since the KCBS inequality uses only sequences of two
measurements, as the CHSH inequality (\ref{chshineq}). A generalization of
Eq.~(\ref{chshwithnoise3}) is
\begin{align}
& \mean{\XX_{\rm KCBS}} := \mean{A_1 B_2} + \mean{C_1 B_2} +
\mean{C_1 D_2} + \mean{E_1 D_2}
\nonumber
\\
&+ \mean{E_1 A_2}
\geq -3 -
( \varepsilon_{AB}+\varepsilon_{CB}+\varepsilon_{CD}+\varepsilon_{ED}+\varepsilon_{EA}).
\label{kbcswithnoise3}
\end{align}
Generalizations of Eqs.~(\ref{chshwithnoise2}) and (\ref{stochcontext}) can also be 
written down in a similar manner.

Also for the KS inequality (\ref{mpineq}), one can deduce
generalizations, which exclude certain types of HV models. The
main problem here is to estimate a term like $\mean{A_1 B_2
C_3}.$ First, an inequality corresponding to Eq.~(\ref{bound1})
is
\begin{align}
p[(A_1^+| & A_1) \mbox{ and }(B_1^+|B_1) \mbox{ and } (C_1^+| C_1)]
\nonumber
\\
\leq &
\;\;
p[A_1^+, B_2^+, C_3^+| A_1 B_2 C_3] +
\nonumber
\\
 &
+
p[(B_1^+|B_1) \mbox{ and } (B_2^-|A_1 B_2)]
\nonumber
\\
 &
+
p[(C_1^+|C_1) \mbox{ and } (C_3^-|A_1 B_2 C_3)],
\label{bound11}
\end{align}
which holds again for any HV model. Then, a direct calculation gives that
one has
\begin{align}
\mean{ABC} &\leq \mean{A_1 B_2 C_3} + 4 p^{\rm flip}[AB]
+ 4 p^{\rm flip}[(AB)C]
\nonumber
\\
\mean{ABC} &\geq \mean{A_1 B_2 C_3} - 4 p^{\rm flip}[AB] - 4 p^{\rm flip}[(AB)C],
\end{align}
where
\begin{align}
p^{\rm flip}&[(AB)C] =
p[(C_1^+|C_1) \mbox{ and }(C_3^-|A_1 B_2C_3)]
\nonumber
\\
&
+ p[(C_1^-|C_1) \mbox{ and } (C_3^+|A_1 B_2 C_3)].
\end{align}
Given these bounds, one arrives at testable inequalities, provided assumptions
on the HV model are made as in the three approaches above. If Eq.~(\ref{assumption})
is assumed, one can directly estimate
$p^{\rm flip}[AB] \leq p^{\rm err}[A_1 B_2]$
and
\begin{align}
 p^{\rm flip}&[(AB)C] \leq p^{\rm err}[(AB)C]
\\
&= p[C_1^+ C_4^-|C_1 A_2 B_3 C_4] + p[C_1^- C_4^+|C_1 A_2 B_3 C_4].
\nonumber
\end{align}
Then, if one writes down the generalized form of
Eq.~(\ref{mpineq}), then there are more correction terms than
in Eq.~(\ref{chshwithnoise2}). Moreover, they involve
sequences of length four. On average, these $p^{\rm err}$ terms
have to be smaller than $2/48 \approx 0.0417$ in order to allow
a violation. Consequently, an experimental test is very
demanding (see also the discussion in subsection
\ref{explimitations}). Finally, generalizations in the sense of
Eqs.~(\ref{stochcontext} and \ref{chshwithnoise3}) can also be
derived in a similar manner.


\section{Experimental implementation}
\label{Sec4}


Experimental tests of noncontextual HV theories have been
carried out with photons \cite{MWZ00, HLZPG03, ARBC09, HLBBR06,
BKSSCRH09}, neutrons \cite{HLBBR06, BKSSCRH09}, laser-cooled
trapped ions \cite{KZGKGCBR09}, and liquid-state nuclear
magnetic resonance systems \cite{MRCL09}. 
In the experiments with photons and neutrons, single particles 
were prepared and measured in a four-dimensional state space 
composed of two two-dimensional state spaces describing the particle's
polarization and the path it was following. In contrast, in a
recent experiment with trapped ions \cite{KZGKGCBR09}, a
composite system  comprised of two trapped ions prepared
in superpositions of two long-lived internal states was used
for testing the KS theorem. In the following, we will describe
this experiment and present details about the amount of
noncompatibility of the observables implemented.


\subsection{Experimental methods}

\begin{figure}
\includegraphics[width=7.6cm]{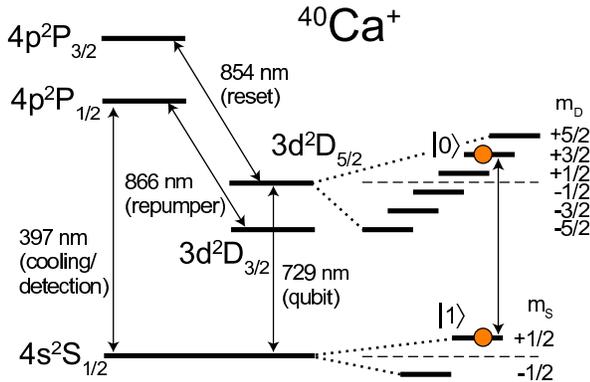}
\caption{\label{fig1:levelscheme} Partial level scheme of $^{40}$Ca$^+$
showing the relevant energy levels and the laser wavelengths needed for
coupling the states. The $D$-states are metastable with a lifetime of
about 1s. A magnetic field of about 4 Gauss is applied to lift the
degeneracy of the Zeeman states. The states $|0\rangle$, $|1\rangle$
used for encoding quantum information are indicated in the figure.}
\end{figure}

Trapped laser-cooled ions are advantageous for these kinds of
measurements because of the highly efficient quantum state preparation
and measurement procedures trapped ions offer. In
Ref.~\cite{KZGKGCBR09}, a pair of $^{40}$Ca$^+$ ions was prepared in a
state space spanned by the states $|00\rangle$, $|01\rangle$,
$|10\rangle$, $|00\rangle$, where $|1\rangle=|S_{1/2},m_S=1/2\rangle$
is encoded in a Zeeman ground state and
$|0\rangle=|D_{5/2},m_D=3/2\rangle$ in a long-lived metastable state
of the ion (see Fig.~\ref{fig1:levelscheme}).


A key element for both preparation and measurement are
laser-induced unitary operations that allow for arbitrary
transformations on the four-dimensional state space. For this,
the entangling operation
$U^{MS}(\theta,\phi)=\exp(-i\frac{\theta}{2}\sigma_\phi
\otimes\sigma_\phi )$ where $\sigma_\phi=\cos(\phi)\,\sigma_x +
\sin(\phi)\,\sigma_y$ is realized by a bichromatic laser field
off-resonantly coupling to transitions involving the ions'
center-of-mass mode along the weakest axis of the trapping
potential \cite{KBZGRB09}. In addition, collective single-qubit
gates
$U(\theta,\phi)=\exp[-i\frac{\theta}{2}(\sigma_\phi\otimes
\eins +\eins \otimes \sigma_\phi)]$ are realized by resonantly
coupling the states $|0\rangle$, $|1\rangle$. Finally, the
single-qubit gate
$U_z(\theta)=\exp(-i\frac{\theta}{2}\sigma_z)$ is implemented
by a strongly focused laser inducing a differential light-shift
on the states of the first ion. This set of operations, ${\cal
S}=\{U_z(\theta),U(\theta,\phi),U^{MS}(\theta,\phi)\}$, which
is sufficient for constructing arbitrary unitary operations,
can be used for preparing the desired input states
$\ket{\psi}$.

A measurement of $\sigma_z$ by a state projection onto the
basis states $|0\rangle$, $|1\rangle$ on one of the ions is
carried out by illuminating the ion with laser light coupling
the $S_{1/2}$ ground state to the short-lived excited state
$P_{1/2}$ and detecting the fluorescence emitted by the ion
with a photomultiplier. Population in $P_{1/2}$ decays back to
$S_{1/2}$ within a few nanoseconds so that thousands of photons
are scattered within a millisecond if the ion was originally in
the state $|1\rangle$. If it is in state $|0\rangle$, it does not
couple to the light field and therefore scatters no photons. In
the experiment, we assign the state $|1\rangle$ to the ion if
more than one photon is registered during a photon collection
period of 250$\mu$s. In this way, the observables $\sigma_z
\otimes \openone$ and $\openone \otimes \sigma_z$ can be
measured.

To measure further observables like $\sigma_i \otimes
\openone$, $\openone \otimes \sigma_j$, or $\sigma_i \otimes
\sigma_j$, the quantum state $\rho$ to be measured is
transformed into $U\rho U^\dagger$ by a suitable unitary
transformation $U$ prior to the state detection. Measuring the
value of $\sigma_z\otimes\openone$ on the transformed state is
equivalent to measuring the observable
$A=U^\dagger(\sigma_z\otimes\openone)U$ on the original state
$\rho$. The measurement is completed by applying the inverse
operation $U^\dagger$ after the fluorescence measurement. The
purpose of this last step is to map the projected state onto an
eigenstate of the observable $A$. In this way, any observable
$A$ with two pairs of degenerate eigenvalues can be measured.
The complete measurement, consisting of unitary transformation,
fluorescence detection and back transformation, constitutes a
quantum nondemolition measurement of $A$. Each measurement of a
quantum state yields one bit of information which carries no
information about other compatible observables.


\subsection{Measurement results}

\begin{figure}
\includegraphics[width=0.6 \columnwidth]{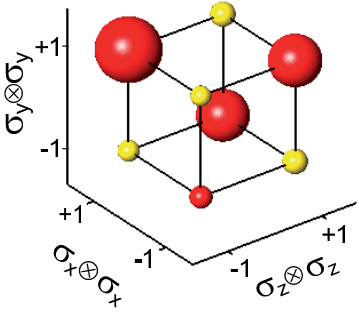}
\caption{\label{fig2:correlations} Measurement correlations for
a sequence of measurements $A_1B_2C_3$ with $A_1=\sigma_z\otimes\sigma_z$,
$B_2=\sigma_x\otimes\sigma_x$, and $C_3=\sigma_y\otimes\sigma_y$ for
a partially entangled input state. The colors indicates whether
$v_1v_2v_3=+1$ (yellow spheres) or $v_1v_2v_3=-1$ (red spheres).
The volume of a sphere is proportional to the likelihood of
finding the corresponding measurement outcome $(v_1,v_2,v_3)$.}
\end{figure}

The measurement procedure outlined above is very flexible and
can be used to consecutively measure several observables on a
single quantum system as illustrated by the following example.
To test inequality Eq.~(\ref{ksinequalityq}) for the
observables of the Mermin-Peres square~(\ref{mpsquare}), the
quantum state
$\ket{\psi}=|11\rangle/{\sqrt{2}}+e^{i\frac{\pi}{4}}(|01\rangle+|10\rangle)/2$
is prepared by a applying the sequence of gates
$U^{MS}(-\pi/2,\pi/4)U^{MS}(-\pi/2,0)U(\pi/2,0)$ to the initial
state $|11\rangle$. The correlations that are found for a
sequence of measurements $A_1B_2C_3$, where
$A_1=\sigma_z\otimes\sigma_z$, $B_2=\sigma_x\otimes\sigma_x$,
and $C_3=\sigma_y\otimes\sigma_y$ are shown in
Fig.~\ref{fig2:correlations}. For this measurement, 1100 copies
of the state were created and measured. Each corner of the
sphere corresponds to a measurement outcome $(v_1,v_2,v_3)$
where $v_k=\pm 1$ is the measurement result for the k$^{th}$
observable. The relative frequencies of the measurement
outcomes are indicated by the volume of the spheres attached to
the corners, and the colors indicates whether $v_1v_2v_3=+1$ or
$v_1v_2v_3=-1$. For perfect state preparation and measurements,
one would expect to observe always $v_1v_2v_3=-1$. Due to
experimental imperfections, the experiment yields $\langle
v_1v_2v_3\rangle=-0.84(2)$. Nevertheless, the experimental
results nicely illustrate the quantum measurement process: the
first measurement gives
$\langle\sigma_z\otimes\sigma_z\rangle=0.00(2)$, i.e. the state
$\ket{\psi}$ is equally likely to be projected onto
$\ket{\Psi_+}=|11\rangle$ ($v_1=+1$) and onto
$\ket{\Psi_-}=(|01\rangle+|10\rangle)/{\sqrt{2}}$ ($v_1=-1$).
In the latter case, the {projected} state $\ket{\Psi_-}$ is
an eigenstate of $\sigma_x\otimes\sigma_x$ and
$\sigma_y\otimes\sigma_y$ so that these measurements give
definite results $v_2=+1$ and $v_3=+1$ (upper left corner of
Fig.~\ref{fig2:correlations}). In the former case, the
{projected} state is not an eigenstate of
$\sigma_x\otimes\sigma_x$ and $v_2=+1$ and $v_3=-1$ are found
with equal likelihood. In this case, $v_2$ and $v_3$ are random
but correlated with $v_2v_3=1$ (the other two strongly
populated corners of Fig.~\ref{fig2:correlations}).


In Ref.~\cite{KZGKGCBR09}, also the other rows and columns of
the Mermin-Peres square (\ref{mpsquare}) were measured for the
state $\ket{\psi}$, and a violation of Eq.~(\ref{ksinequalityq}) was found
with $\mean{\chi_{\rm KS}}=5.36(4)$. Also different input states
were investigated to check that the violation is indeed
state-independent. The fact that the result falls short of the
quantum mechanical prediction of $\mean{\chi_{\rm KS}}=6$ is due
to imperfections in the measurement procedure. These
imperfections could be incorrect unitary transformations, but
also errors occurring during the fluorescence measurement.

An instructive test consists in repeatedly measuring the
same observable on a single quantum system and analyzing
the measurement correlations. Table~\ref{Tab:zIcorrelations}
shows the results of five consecutive measurements of
$A=\sigma_z\otimes\openone$ on a maximally mixed state
based on 1100 experimental repetitions.


\begin{table}
\caption{Measurement correlations $\langle A_iA_{j}|A_1\ldots A_5\rangle$
between repeated measurements of $A=\sigma_z\otimes\openone$ for a maximally
mixed state. Observing a correlation of $\langle A_iA_{j}|A_1\ldots A_5\rangle = \alpha_{ij}$
means that the probability for the measurement results of $A_i$ and $A_j$ to
coincide equals $(\alpha_{ij}+1)/2.$
}
\centering
\begin{tabular}{c|ccccc}
\hline\hline
Measurement & 2         &        3  & 4       & 5\\
\hline
1           & 0.97(1)   & 0.97(1)   & 0.96(1) & 0.95(1)\\
2           &           & 0.97(1)   & 0.97(1) & 0.96(1)\\
3           &           &           & 0.98(1) & 0.98(1)\\
4           &           &           &         & 0.98(1)\\
\hline
\end{tabular}
\label{Tab:zIcorrelations}
\end{table}


\begin{table}
\caption{Measurement correlations $\langle A_iA_{j}|A_1\ldots A_5\rangle$
between repeated measurements of $A=\sigma_x\otimes\sigma_x$
for a maximally mixed state.
Observing a correlation of $\langle A_iA_{j}|A_1\ldots A_5\rangle = \alpha_{ij}$
means that the probability for the measurement results of $A_i$ and $A_j$ to
coincide equals $(\alpha_{ij}+1)/2.$}
\centering
\begin{tabular}{c|ccccc}
\hline\hline
Measurement & 2         &        3  & 4       & 5\\
\hline
1           & 0.94(1)   & 0.88(1)   & 0.82(2) & 0.80(2)\\
2           &           & 0.93(1)   & 0.87(2) & 0.84(2)\\
3           &           &           & 0.90(1) & 0.87(2)\\
4           &           &           &         & 0.93(1)\\
\hline
\end{tabular}
\label{Tab:xxcorrelations}
\end{table}

As expected, the correlations $\langle A_iA_{i+k}|A_1\ldots
A_5\rangle$ are independent of the measurement number $i$
within the error bars. However, the correlations become smaller
and smaller the bigger $k$ gets. Table~\ref{Tab:xxcorrelations}
shows another set of measurements correlations $\langle
A_iA_{j}|A_1\ldots A_5\rangle$, where
$A=\sigma_x\otimes\sigma_x$. Here, the correlations are
slightly smaller, since entangling interactions are needed for
mapping $A$ onto $\sigma_z\otimes\openone$, which is
experimentally the most demanding step.

It is also interesting to compare the correlations $\langle
A_1A_3|A_1A_2A_3\rangle$ with the correlations $\langle
A_1A_3|A_1B_2A_3\rangle$ for an observable $B$ that is
compatible with $A$. For $A=\sigma_x\otimes\sigma_x$ and
$B=\sigma_z\otimes\sigma_z$, we find $\langle
A_1A_3|A_1A_2A_3\rangle=0.88(1)$ and $\langle
A_1A_3|A_1B_2A_3\rangle=0.83(2)$ when measuring a maximally
mixed state;
i.e., it seems that the intermediate measurement of $B$
perturbs the correlations slightly more than an intermediate
measurement of $A$. Similar results are found for a singlet
state, where $\langle A_1A_3|A_1A_2A_3\rangle=0.92(1)$,
$\langle B_1B_3|B_1B_2B_3\rangle=0.91(1)$, but $\langle
A_1A_3|A_1B_2A_3\rangle=0.90(1)$, and $\langle
B_1B_3|B_1A_2B_3\rangle=0.89(1)$.
Because of $\langle B_1B_3|B_1A_2B_3\rangle=1-2p^{\rm
err}(B_1A_2B_3)$, correlations of the type $\langle
B_1B_3|B_1A_2B_3\rangle$ are required for checking  inequality
(\ref{chshwithnoise2}) that takes into account disturbed HVs.


\subsection{Experimental limitations}
\label{explimitations}

There are a number of error sources contributing to
imperfect state correlations, the most important being:

(i) {\it Wrong state assignment based on fluorescence data.}
During the 250~$\mu$s detection period of the current
experiment, the number of detected photons has a Poissonian
distribution with an average number of
$\overline{n}_{|1\rangle}=8$ photons if the ion is in state
$|1\rangle$. If the ion is in state $|0\rangle$, it does not
scatter any light, however, light scattered from trap electrodes
gives rise to a Poissonian distribution with an average number
of $\overline{n}_{|0\rangle}=0.08$ photons. These photon count
distributions slightly overlap. The probability of detecting 0
or 1 photons even though the ion is in the bright state, is
0.3\%. The probability of detecting more than one photon if the
ion is in the dark state is also 0.3\%. Therefore, if the
threshold for discriminating between the dark and the bright
state is set between 1 and 2, the probability for wrongly
assigning the quantum state is $0.3\%$. Making the detection
period longer would reduce this error but increase errors
related to decoherence of the other ion's quantum state that is
not measured.

(ii) {\it Imperfect optical pumping.} During fluorescence detection,
the ion leaves the computational subspace $\{|0\rangle,|1\rangle\}$
if it was in state $|1\rangle$ and can also populate the state
$|S_{1/2},m_S=-1/2\rangle$. To prevent this leakage, the ion is
briefly pumped on the $S_{1/2}\leftrightarrow P_{1/2}$ transition
with $\sigma_+$-circularly polarized light to pump the population
back to $|1\rangle$. Due to imperfectly set polarization and
misalignment of the pumping beam with the quantization axis,
this pumping step fails with a probability of about 0.5\%.

(iii) {\it Interactions with the environment.} Due to the
non-zero differential Zeeman shift of the states used for
storing quantum information, superposition states dephase in
the presence of slowly fluctuating magnetic fields. In
particular, while measuring one ion by fluorescence detection,
quantum information stored in the other ion dephases. We
partially compensate for this effect by spin-echo-like
techniques \cite{HRB08} that are based on a transient storage
of superposition states in a pair of states having an opposite
differential Zeeman shift as compared to the states $|0\rangle$
and $|1\rangle$. A second interaction to be taken into account
is spontaneous decay of the metastable state $|0\rangle$ which
however only contributes an error of smaller than $0.1\%$.

(iv) {\it Imperfect unitary operations.} The mapping operations
are not error-free. This concerns in particular the entangling
gate operations needed for mapping the eigenstate subspace of a
spin correlation $\sigma_i\otimes\sigma_j$ onto the
corresponding subspaces of $\sigma_z\otimes\openone$. For this
purpose, a M{\o}lmer-S{\o}rensen gate operation
$U^{MS}(\pi/2,\phi)$ \cite{SM99, KBZGRB09} is used. This gate
operation has the crucial property of requiring the ions only
to be cooled into the Lamb-Dicke regime. In the experiments,
the center-of-mass mode used for mediating the gate interaction
is in a thermal state with an average of 18 vibrational quanta.
In this regime, the gate operation is capable of mapping
$|11\rangle$ onto a state $|00\rangle+e^{i\phi}|11\rangle$ with
a fidelity of about 98\%. Taking this fidelity as being
indicative of the gate fidelity, one might expect errors of
about 4\% in each measurement of spin correlations
$\sigma_i\otimes\sigma_j$ as the gate is carried out twice,
once before and once after the fluorescence measurement.

These error sources prevented us from testing a generalization
of inequality (\ref{mpineq}) as discussed in subsection
\ref{subsec:KSineqnonperf}. Measurement of the correlations
$\langle B_1B_3|B_1A_2B_3\rangle$ and $\langle
C_1C_4|C_1A_2B_3C_4\rangle$ resulted in error terms $p^{\rm
err}$ that were about $0.06$ for sequences involving three
measurements and about $0.1$ for sequences with four
measurements, i.~e. twice as big as required for observing a
violation of (\ref{mpineq}). However, the experimental errors
were small enough to demonstrate a violation of the
CHSH-like inequality (\ref{chshwithnoise2}), valid for
nonperfectly compatible observables \cite{KZGKGCBR09}. A test
of the inequality (\ref{stochcontext}) would become possible if
the error rates could be further reduced.


\section{Contextual HV models}
\label{Sec5}


In this section we will introduce two HV models which are contextual
in the sense of Eq.~(\ref{ncdef}) and violate the inequalities discussed 
in Sec. \ref{Sec2}. We first discuss a simple model which violates
inequality~(\ref{chshwithnoise2}), and then a more complex one,
which reproduces all measurement results for a (finite-dimensional)
quantum mechanical system. These models are useful to point out
which counterintuitive properties a HV model must have to reproduce
the quantum predictions, and which further experiments can rule out
even these models.


\subsection{A simple HV model leading to a violation of
inequality~(\ref{chshwithnoise2})}


We will show here that violation of inequality~(\ref{chshwithnoise2})
can be achieved simply by allowing the HV model to remember what
measurements have been performed and what the outcome was. The basic
idea of the model is very simple (cf.\ the more complicated
presentation in \cite{La Cour09a}).

The task is to construct a simple HV model for our four dichotomic
observables $A,B,C$, and $D$. The HV $\lambda$ is taken to be a
quadruple with entries taken from the set $\{+,-,\oplus,\ominus\}$,
the latter two cases will be called ``locked'' in what follows,
signifying that the value is unchanged whenever a compatible
measurement is made. For convenience, we can write
$\lambda=(A^+,B^+,C^+,D^+)$ or $\lambda=(A^+,B^-,C^\oplus,D^\ominus)$,
etc, and we take the initial distribution to be probability $1/2$ of
either $(A^+,B^+,C^+,D^+)$ or $(A^-,B^-,C^-,D^-)$.  The measurement of
an observable is simply reporting the appropriate sign, and locking
the value in the position. To make the model contextual, we add the
following mechanism:
\begin{description}
\item (a) If $A$ is measured, then the sign of $D$ is reversed and
  locked unless it is locked.
\item (b) If $D$ is measured, then the sign of $A$ is reversed and
  locked unless it is locked.
\end{description}
For the case $\lambda=(A^+,B^+,C^+,D^+)$, the measurement results
when measuring inequality~(\ref{chshwithnoise2}) will be as follows.
\begin{description}
\item (i) Measurement of $A_1$ will yield $A_1^+$ and
  $\lambda=(A^\oplus,B^+,C^+,D^\ominus)$, and for the next measurement
  one obtains $B_2^+$ or $D_2^-$.
\item (ii) Measurement of $B_1$ will yield $B_1^+$ and
  $\lambda=(A^+,B^\oplus,C^+,D^+)$, and further one obtains
  $A_2^+B_3^+$ or $C_2^+B_3^+$.
\item (iii) Measurement of $C_1$ will yield $C_1^+$ and
  $\lambda=(A^+,B^+,C^\oplus, D^+)$, and we'll obtain $B_2^+$ or
  $D_2^+$ afterwards.
\item (iv) Measurement of $D_1$ will yield $D_1^+$ and
  $\lambda=(A^\ominus,B^+,C^+,D^\oplus)$, and we'll obtain
  $C_2^+D_3^+$ or $A_2^-D_3^+$. The last is because a measurement of
  $A_2$ will not change $D^\oplus$ since it is locked.  In this case,
  after a measurement of $A_2$ the HVs are $\lambda=(A^\ominus,
  B^+,C^+,D^\oplus)$.
\end{description}
The case $\lambda=(A^-,B^-,C^-,D^-)$ is the same with reversed
signs. This means that
\begin{equation}
\mean{A_1 B_2} = \mean{C_1 B_2} = \mean{C_1 D_2} = -\mean{A_1 D_2}=1,
\end{equation}
and
\begin{equation}
  \begin{split}
    p^{\rm err}[&B_1 A_2 B_3] = p^{\rm err}[B_1 C_2 B_3] = p^{\rm
      err}[D_1 C_2 D_3]\\
    &= p^{\rm err}[D_1 A_2 D_3]=0.
  \end{split}
\end{equation}
Hence, this model leads to the maximal violation of
Eq.~(\ref{chshwithnoise2}).

In this model, the observables $A$ and $D$ are compatible in the sense
of Definition 1, but they maximally violate the noncontextuality
condition in Eq.~(\ref{ncdef}). It is easy to verify that $p^{\rm
  flip}[AD]=1$, so that the assumption (\ref{assumption}) does not
hold. We argue that in this model, the change in the outcome $D$
cannot be explained as merely due to a disturbance of the system from
the experimental procedures when measuring $A_1$. It should therefore
be no surprise that the inequality~(\ref{chshwithnoise2}) is violated
by the model. Finally, note that a model behavior like this would
create problems in any argument to establish noncontextuality via
repeatability of compatible measurements, even for joint measurements
as discussed in Section \ref{Sec2}A, and not only in the sequential
setting used here.

\subsection{A HV model explaining all quantum mechanical predictions}


Let us now introduce a detailed HV model which reproduces all
the quantum predictions for sequences of measurements. In
a nutshell, this contextual HV model is a translation of a
machine that classically simulates a quantum system.

We consider the case that only dichotomic measurements are
performed on the quantum mechanical system. Therefore, any
observable $A$ decomposes into $A=\Pi^A_+-\Pi^A_-$ with
orthogonal projectors $\Pi_+$ and $\Pi_-$. For a mixed state
$\vr$, a measurement of this observable produces the result
$+1$ with probability $ p(A^+)=\trace(\Pi^A_+ \rho)$, and the
result $-1$ with probability $ p(A^-)=\trace(\Pi^A_- \rho)$. In
addition, the measurement apparatus will modify the quantum
state according to
\begin{equation}
\label{e756}
\vr \mapsto \frac{\Pi^A_\pm \vr \Pi^A_\pm}{\trace(\Pi^A_\pm \vr)},
\end{equation}
depending on the measurement result $\pm 1$.

This behavior can exactly be mimicked by a HV model, if we allow the
value of the HV to be modified by the action of the measurement. If
$\HH$ is the Hilbert space of the quantum system, we use two types
of HVs. First, we use parameters $0\le \lambda^A<1$, $0\le
\lambda^B<1$, etc.~for each observable $A$, $B$, etc.~and second we
use a normalized vector $\ket{\psi}\in \HH$.

Then, for given values of all these parameters, we associate to
any observable the measurement result as follows: We define
$q^A=\bra\psi\Pi^A_-\ket\psi$ and let the model predict the
measurement result: $-1$ if $\lambda^A<q^A$, and $+1$ if $q^A
\leq \lambda^A.$ Furthermore, depending on the measurement
result, the values of the HVs $\lambda^A$ and $\ket\psi$ change
according to
\begin{equation}
\label{e17077}
\lambda^A
\mapsto
\left\{
\begin{split}
&\frac{\lambda^A}{q^A} &\text{if $\lambda^A<q^A$},\\
&\frac{\lambda^A-q^A}{1-q^A} & \text{if $\lambda^A \geq q^A$},
 \end{split}
\right.
\end{equation}
and
\begin{equation}
\label{e8753}
 \ket\psi \mapsto
\left\{
 \begin{split}
& \frac{\Pi^A_-\ket\psi}{\sqrt{q^A}} & \text{if $\lambda^A<q^A$},\\
& \frac{\Pi^A_+\ket\psi}{\sqrt{1-q^A}} & \text{if $\lambda^A \geq
q^A$}.
 \end{split}
\right.
\end{equation}

Let us now fix the initial probability distribution of the HVs.
The experimentally accessible probability distributions
$p(\lambda^A,\lambda^B,\dotsc;\psi)$ shall not depend on the
parameters $\lambda^A$, $\lambda^B,\dotsc,$ that is,
$p(\lambda^A,\lambda^B,\dotsc;\psi)=
p({\lambda'}^A,{\lambda'}^B,\dotsc;\psi)$. Hence we write
$p(\psi)=\int \! d\lambda^A d \lambda^B \dotsm
p(\lambda^A,\dotsc;\psi)$. The probability distribution
$p(\psi)$ and the measure $d\psi$ are chosen such that
\begin{equation}
 \vr_p = \int \!\!d\psi\; p(\psi) \ket\psi\!\bra\psi\,
\end{equation}
is the corresponding quantum state.

We now verify that this model indeed reproduces the quantum
predictions. If the initial distribution is $p$, then the
probability to obtain the result $-1$ for $A$ is given by
 \bea
 p^A_-&= & \int_{\lambda^A<q^A}\!\!d\lambda^A d\psi\; p(\psi)
= \int\!\!d\psi\; \bra\psi \Pi^A_- \ket\psi\, p(\psi) \nonumber
\\
 \vphantom{\frac{1}{2}}&=& \trace(\rho_p \Pi^A_-),
 \eea
and hence is in agreement with the quantum prediction. Due to
the transformations in Eq.~\eqref{e17077} and
Eq.~\eqref{e8753}, the probability distribution changes by the
action of the measurement, $p\mapsto p'$. The new distribution
$p'$ again does not depend on $\lambda^A$ and, in case of the
measurement result $-1$, we have
\begin{equation}
p'(\psi)= \frac{1}{p^A_-}\int\!\! d\psi' {q'}^A\;
\delta\left(\ket\psi- \frac{\Pi^A_-\ket{\psi'}}{\sqrt{{q'}^A}}\right)\,
p(\psi'),
\end{equation}
where $\delta$ denotes Dirac's $\delta$-distribution and
${q'}^A=\bra{\psi'} \Pi_-^A \ket{\psi'}$. The new corresponding
mixed state is given by
 \bea
 \vr_{p'} &= &\int \!\! d\psi\;
 p'(\psi) \ket\psi\!\bra\psi \nonumber
\\
 &= & \frac{1}{p^A_-} \int\!\! d\psi'\; p(\psi')
 \Pi^A_-\ket{\psi'}\!\bra{\psi'}\Pi^A_- \nonumber
\\
 &= & \frac{\Pi_-^A \vr_p \Pi_-^A}{\trace(\vr_p \Pi_-^A)}.
 \eea
This demonstrates that the transformation in Eq.~\eqref{e756}
is suitably reproduced by $\vr_p \mapsto \vr_{p'}$. An
analogous calculation can be performed for the measurement
result $+1$.

Let us illustrate that this model is actually contextual, as
defined in Eq.~\eqref{ncdef}. As an example, we choose two
commuting observables $A=\Pi^A_+-\Pi^A_-$ and
$B=\Pi^B_+-\Pi^B_-$ with the property that, for some pure state
$\ket\psi$, we have $\bra\psi A_1 B_2 \ket\psi=+1$, while
$\bra\psi B \ket\psi < 1$. An example would be
$A=\sigma_z\otimes\openone$ and $B=-\openone\otimes \sigma_z$
with $\ket\psi$ being the singlet state. Then, after a
measurement of $A_1$, the result of a subsequent measurement of
$B_2$ is fixed and hence independent of $\lambda^B$. However,
if $B$ is measured without a preceding measurement of $A$, then
the result of $B$ will be $-1$ if $\lambda^B< \bra\psi \Pi^B_-
\ket\psi$, and $+1$ else. Hence, in our particular model, given
the preparation of $\ket\psi$, $v(B_1)$ depends on $\lambda^B$,
while $v(B_2|A_1B_2)$ only depends on $\lambda^A$. However, the
model does not allow special correlations between $\lambda^A$
and $\lambda^B$ and hence the model is contextual, i.e.,
necessarily there are experimentally accessible values of the
HVs, such that Eq.~\eqref{ncdef} is violated.


\section{Conclusions}
\label{Sec6}


Experimental quantum contextuality is a potential source of new 
applications in quantum information processing, and a chance 
to expand our knowledge on the reasons why quantum resources 
outperform classical ones. In some sense, experimental 
quantum contextuality is an old discipline, since most Bell 
experiments are just experiments ruling out noncontextual HV models, 
since they do not fulfill the required spacelike separation needed 
to invoke locality as a physical motivation behind the assumption of
noncontextuality. The possibility of observing state-independent quantum 
contextuality, however, is a recent development. It shows that the power of QM is 
not necessarily in some particular states, but also in some sets of 
measurements which can reveal nonclassical behavior of any quantum
state.

These experiments must satisfy some requirements which are not 
explicitly  needed for tests of Bell inequalities. An important 
requirement is that one has to test experimentally, to which extent
the implemented measurements are indeed compatible. In this paper, 
we have discussed how to deal with the inevitable errors, preventing
us from implementing perfectly compatible measurements. The problem 
of not-perfectly compatible observables is not fatal, but should be 
taken into account with care. 

We have presented three approaches how additional requirements can be 
used to exclude the possibility of noncontextual explanations 
of the experimental results, and we have applied them to three specific 
inequalities of particular interest: a CHSH-like noncontextuality 
inequality using sequential measurements on individual systems, 
which can be violated by specific states of four or more levels, 
a KCBS noncontextuality inequality using sequential measurements 
on individual systems, which can be violated by specific states 
of three or more levels, and a KS inequality coming from the 
Mermin-Peres square which is violated 
by any state of a four-level system. Similar methods can be applied 
to any noncontextuality inequality, irrespective of the number 
of sequential measurements or the dimensionality of the Hilbert
space.

The main motivation was to provide experimentalists with 
inequalities to rule out noncontextual HV models unambiguously, 
if some additional assumptions are made. We have shown that a 
recent experiment with trapped ions already ruled out some of 
these HV models. By providing examples of HV models, we have 
seen that these extra 
assumptions are not necessarily satisfied by very artificial HV 
models. Nevertheless they lead to natural extensions of the assumption 
of noncontextuality, and allow us to reach conclusions about HV models 
in realistic experiments with nonperfect devices.  
An interesting line of future research will be to investigate how these 
extra assumptions can be replaced by fundamental physical principles
such as locality in experiments where the system under observation 
is entangled with a distant system on which additional measurements 
can be performed.


\section*{Acknowledgments}

The authors thank R.~Blatt, J.~Emerson, B.R.~La~Cour, O.~Moussa, 
and R.W.~Spekkens for discussions 
and acknowledge support by the Austrian Science Fund (FWF), 
the European Commission (SCALA, OLAQUI and QICS networks  and the 
Marie-Curie program), the Institut f\"ur Quanteninformation GmbH, 
the Spanish MCI Project No.~FIS2008-05596, and the Junta de 
Andaluc\'{\i}a Excellence Project No.~P06-FQM-02243. 
A.C.~and J.-\AA.\ L.~thank the IQOQI for its hospitality. 
This material is based upon work supported in part by IARPA.

\section*{Appendix}

In Sec.~\ref{def_of_compatibility} we discussed the notion of
compatibility for subsequent measurements. In this Appendix we
provide two examples which demonstrate that both parts of
Definition~1 are independent. We then show that the statement
of compatibility can be simplified to involve sequences of
length~2 only.

\paragraph{Mutual independence of Definition~1 (i) and Definition~1 (ii).}
For an example that (i) does not include (ii), assume that the
expectation value of $A$ depends on whether the first measurement in the
sequence is $A$ or $B$. Then $\mean{A_1|A_1B_2}\ne \mean{A_2|B_1A_2}$
and hence condition (ii) is violated. However, such a model is not in
conflict with condition (i), if once $A$ was measured, the value of $A$
stays unchanged for the rest of the sequence.

For the converse, assume a HV model where the expected value
$\mean A$ does not depend on the results of any previous
measurement. Then, for any sequence and any $k$, $\mean A=
\mean{A_k|S}$ and, hence, condition (ii) is satisfied. However,
$p(A^+_1A^-_2|A_2A_2)>0$, unless $\mean{A_1 A_2}=1$, and thus
condition (i) is violated.

\paragraph{Compatibility for sequences of length~2.}
Assume that, for any preparation procedure, $A$ and $B$ obey
\begin{equation}\label{e10110}
\mean{A_1}= \mean{A_2|A_1A_2}= \mean{A_2|B_1A_2},
\end{equation}
i.e., condition (ii) of Definition~1 is satisfied for sequences
of length~2. Then, for a sequence $S$ of length $k$ we have
either $S= S'B$ or $S= S'A$, where $S'$ is a sequence of length
$k-1$. In a measurement of $S$, we can consider $S'$ to be part
of the preparation procedure and then apply Eq.~\eqref{e10110}.
It follows that $\mean{A_{k+1}|SA}= \mean{A_k|S'A}$ and
eventually $\mean{A_{k+1}|SA}= \mean{A_1}$ by induction.

In a similar fashion we reduce condition (i) of Definition~1
for dichotomic observables. For an experimentally accessible probability distribution
$p_\mathrm{exp} (\lambda)$, we denote by $\tilde
p_\mathrm{exp}(\lambda)$ the distribution obtained by a
measurement of $A$ and a postselection of the result $+1$.
Then, for a sequence $S$ of length $k$,
\begin{align}
 p(A_1^+ A_{k+2}^-|ASA)
   &= \tilde{p}(A_{k+1}^-|SA)\, p(A_1^+|A_1) \nonumber \\
   &= \tilde{p}(A_1^-|A)\, p(A_1^+|A)\nonumber \\
   &= p(A_1^+ A_2^-|A_1A_2),
\end{align}
where for the second equality we used that $\mean{A_{k+1}|SA}=
\mean{A_1}$ holds for $\tilde p$. It follows that a set of
dichotomic observables $\Xi$ is compatible if and only if, for any
preparation and any $A,B\in \Xi$, $\mean{A_1A_2}= 1$ and
Eq.~\eqref{e10110} holds. In particular, this proves the
assertion that pairwise compatibility of three or more
observables is equivalent to an extended definition of
compatibility involving sequences of all compatible
observables.



\begin{thebibliography}{99}


\bibitem{VonNeumann31}
J. von Neumann,
 Ann. of Math. {\bf 32}, 191 (1931).

\bibitem{EPR35}
 A. Einstein, B. Podolsky, and N. Rosen,
 Phys.~Rev. {\bf 47}, 777 (1935).

\bibitem{Bohr35}
 N. Bohr,
 Phys. Rev. {\bf 48}, 696 (1935).


\bibitem{WW01}
 R. F. Werner and M. M. Wolf,
 Quantum Inf. Comput. {\bf 1}(3), 1 (2001).

\bibitem{BH93}
 D. Bohm and B. J. Hiley,
 {\em The Undivided Universe. An Ontological Interpretation of
 Quantum Theory}
 (Routledge, London, 1993).

\bibitem{Holland93}
 P. R. Holland,
 {\it The Quantum Theory of Motion.
 An Account of the de Broglie-Bohm Causal
 Interpretation of Quantum Mechanics}
 (Cambridge University Press, Cambridge, UK, 1993).


\bibitem{Bell64}
 J. S. Bell,
 Physics (Long Island City, NY) {\bf 1}, 195 (1964).

\bibitem{CHSH69}
 J. F. Clauser, M. A. Horne, A. Shimony, and R. A. Holt,
 Phys. Rev. Lett. {\bf 23}, 880 (1969),
{\it ibid.} {\bf 24}, 549 (1970).

\bibitem{ADR82}
 A. Aspect, J. Dalibard, and G. Roger,
 Phys. Rev. Lett. {\bf 49}, 1804 (1982).

\bibitem{WJSWZ98}
 G. Weihs, T. Jennewein, C. Simon, H. Weinfurter, and
 A. Zeilinger,
 Phys. Rev. Lett. {\bf 81}, 5039 (1998).

\bibitem{RKMSIMW01}
 M. A. Rowe, D. Kielpinski, V. Meyer, C. A. Sackett, W. M.
 Itano, C. Monroe, and D. J. Wineland,
 Nature (London) {\bf 409}, 791 (2001).

\bibitem{MMMOM08}
 D. N. Matsukevich, P. Maunz, D. L. Moehring, S. Olmschenk, and
 C. Monroe,
 Phys. Rev. Lett. {\bf 100}, 150404 (2008).

\bibitem{RWVHKCZW09}
W. Rosenfeld, M. Weber, J. Volz, F. Henkel, M. Krug,
 A. Cabello, M. \.{Z}ukowski, and H. Weinfurter,
 Adv. Sci. Lett. {\bf 2}, 469 (2009).


\bibitem{Leggett03}
 A. J. Leggett,
 Found. Phys. {\bf 33}, 1469 (2003).

\bibitem{GPKBZAZ07}
 S. Gr\"oblacher, T. Paterek, R. Kaltenbaek, \v{C}. Brukner,
 M. \.{Z}ukowski, M. Aspelmeyer, and A. Zeilinger,
 Nature (London) {\bf 446}, 871 (2007); {\em ibid.} {\bf 449},
 252 (2007).

\bibitem{BBGKLLS08}
 C. Branciard, N. Brunner, N. Gisin, C. Kurtsiefer,
 A. Lamas-Linares, A. Ling, and V. Scarani,
 Nat. Phys. {\bf 4}, 681 (2008).


\bibitem{Specker60}
 E. Specker,
 Dialectica {\bf 14}, 239 (1960).

\bibitem{Bell66}
 J. S. Bell,
 Rev. Mod. Phys. {\bf 38}, 447 (1966).

\bibitem{KS67}
 S. Kochen and E. P. Specker,
 J. Math. Mech. {\bf 17}, 59 (1967).


\bibitem{RS93}
 S. M. Roy and V. Singh,
Phys. Rev. A {\bf 48}, 3379 (1993).

\bibitem{CG98}
 A. Cabello and G. Garc\'{\i}a-Alcaine,
 Phys. Rev. Lett. {\bf 80}, 1797 (1998).

\bibitem{SZWZ00}
 C. Simon, M. \.Zukowski, H. Weinfurter, and A. Zeilinger,
 Phys.~Rev.~Lett. {\bf 85}, 1783 (2000).

\bibitem{SBZ01}
 C. Simon, \v{C}. Brukner, and A. Zeilinger,
 Phys.~Rev.~Lett. {\bf 86}, 4427 (2001).

\bibitem{Larsson02}
 J.-\AA. Larsson,
 Europhys. Lett. {\bf 58}, 799 (2002).


\bibitem{Meyer99}
 D. A. Meyer,
Phys. Rev. Lett. {\bf 83}, 3751 (1999).

\bibitem{Kent99}
 A. Kent,
Phys. Rev. Lett. {\bf 83}, 3755 (1999).

\bibitem{Mermin99}
 N. D. Mermin,
 quant-ph/9912081.

\bibitem{CK00}
 R. Clifton and A. Kent,
 Proc. R. Soc. London, Ser. A {\bf 456}, 2101 (2000).

\bibitem{HKSS01}
 H. Havlicek, G. Krenn, J. Summhammer, and K. Svozil,
 J. Phys.~A {\bf 34}, 3071 (2001).

\bibitem{Appleby02}
 D. M. Appleby,
 Phys. Rev.~A {\bf 65}, 022105 (2002).

\bibitem{Cabello02}
 A. Cabello,
 Phys. Rev.~A {\bf 65}, 052101 (2002).

\bibitem{Breuer02a}
 T. Breuer,
 in {\em Non-locality and Modality},
 edited by T. Placek and J. Butterfield
 (Kluwer Academic, Dordrecht, Holland, 2002),
 p.~195.

\bibitem{Breuer02b}
 T. Breuer,
Phys. Rev. Lett. {\bf 88}, 240402 (2002).

\bibitem{BK04}
 J. Barrett and A. Kent,
 Stud. Hist. Phil. Sci. Part B: Stud. Hist. Philos. Mod. Phys. {\bf 35}, 151 (2004).


\bibitem{MWZ00}
 M. Michler, H. Weinfurter, and M. \.{Z}ukowski,
 Phys. Rev. Lett. {\bf 84}, 5457 (2000).

\bibitem{HLZPG03}
 Y.-F. Huang, C.-F. Li, Y.-S. Zhang, J.-W. Pan, and G.-C. Guo,
 Phys. Rev. Lett. {\bf 90}, 250401 (2003).

\bibitem{HLBBR06}
 Y. Hasegawa, R. Loidl, G. Badurek, M. Baron, and H. Rauch,
 Phys.~Rev.~Lett. {\bf 97}, 230401 (2006).


\bibitem{Spekkens05}
 R. W. Spekkens,
 Phys. Rev. A {\bf 71}, 052108 (2005).

\bibitem{SBKTP09}
 R. W. Spekkens, D. H. Buzacott, A. J. Keehn, B. Toner, and G. J. Pryde,
 Phys. Rev. Lett. {\bf 102}, 010401 (2009).


\bibitem{CFRH08}
 A. Cabello, S. Filipp, H. Rauch, and Y. Hasegawa,
 Phys. Rev. Lett. {\bf 100}, 130404 (2008).

\bibitem{KCBS08}
 A. A. Klyachko, M. A. Can, S. Binicio\u{g}lu, and A. S. Shumovsky,
 Phys. Rev. Lett. {\bf 101}, 020403 (2008).

\bibitem{Cabello08}
 A. Cabello,
 Phys. Rev. Lett. {\bf 101}, 210401 (2008).

\bibitem{BBCP09}
 P. Badzi{\c a}g, I. Bengtsson, A. Cabello, and I. Pitowsky,
 Phys. Rev. Lett. {\bf 103}, 050401 (2009).


\bibitem{KZGKGCBR09}
 G. Kirchmair, F. Z\"ahringer, R. Gerritsma, M. Kleinmann,
 O. G{\"u}hne, A. Cabello, R. Blatt, and C.~F. Roos,
 Nature (London) {\bf 460}, 494 (2009).

\bibitem{BKSSCRH09}
 H. Bartosik, J. Klepp, C. Schmitzer, S. Sponar,
 A. Cabello, H. Rauch, and Y. Hasegawa,
 Phys.~Rev.~Lett. {\bf 103}, 040403 (2009).

\bibitem{ARBC09}
E. Amselem, M. R{\aa}dmark, M. Bourennane, and A. Cabello,
 Phys. Rev. Lett. {\bf 103}, 160405 (2009).

\bibitem{LHGSZLG09}
 %
B. H. Liu, Y. F. Huang, Y. X. Gong, F. W. Sun,
 Y. S. Zhang, C. F. Li, and G. C. Guo,
 Phys. Rev. A {\bf 80}, 044101 (2009).

\bibitem{MRCL09}
O. Moussa, C. A. Ryan, D. G. Cory, and R. Laflamme,
 \eprint{arXiv:0912.0485}.


\bibitem{La Cour09a}
 B. R. La Cour,
 Phys. Rev. A {\bf 79}, 012102 (2009).


\bibitem{Bell82}
J. S. Bell,
 Found. Phys. {\bf 12}, 989 (1982).


\bibitem{Peres99}
 A. Peres,
 Found. Phys. {\bf 29}, 589 (1999).


\bibitem{Peres90}
 A. Peres,
 Phys. Lett. A {\bf 151}, 107 (1990).

\bibitem{Mermin90}
 N. D. Mermin,
 Phys. Rev. Lett. {\bf 65}, 3373 (1990).


\bibitem{KBZGRB09}
 G. Kirchmair, J. Benhelm, F. Z\"ahringer, R. Gerritsma, C. F. Roos, and R. Blatt,
 New J. Phys. {\bf 11}, 023002 (2009).

\bibitem{HRB08}
 H.~H\"affner, C.~F.~Roos, and R.~Blatt,
 Phys.~Rep. {\bf 469}, 155 (2008).

\bibitem{SM99}
 A. S{\o}rensen and K. M{\o}lmer,
 Phys. Rev. Lett. {\bf 82}, 1971 (1999).


\end{thebibliography}
\end{document}